# A BLOCKCHAIN-BASED INTELLIGENT RECOMMENDER SYSTEM FRAMEWORK FOR ENHANCING SUPPLY CHAIN RESILIENCE

Yang Hu[1,2*], Davy Monticolo [2], Pezhman Ghadimi[1]

[1]Laboratory for Advanced Manufacturing Simulation and Robotics, School of Mechanical & Materials Engineering, University College Dublin, Belfield, Dublin, Ireland

[2]Université de Lorraine, ERPI, Nancy, France

yahu@tcd.ie davy.moticolo@univ-lorraine.fr pezhman.ghadimi@ucd.ie

**Abstract**

Applying advanced digital technologies such as artificial intelligence (AI), blockchain (BLC), bigdata analytics (BDA) and digital twin (DT)/simulations to enhance supply chain resilience (SCRes) has been widely discussed in light of the global pandemic, regional conflicts, and the technology revolution such as Industry 4.0 and 5.0. Previous studies are limited at the conceptual level as the proactive SCRes measure with a standalone fashion. The intelligent recommendation system (IRS) obtains the capabilities for enhancing SCRes as a reactive digital measure. However, the utilization of the IRS as the SCRes enhancement tool is neglected, investigation on implementing the IRS for the SC disruption response is yet to come. To close these gaps, a data-driven supply chain disruption response IRS baseline framework was proposed by this research as an initial SCRes reactive solution. To guarantee the reliability of the proposed IRS as a stable, secure, and resilient decision support system, blockchain technology is integrated into the baseline architecture. The BLC-IRS framework is demonstrated with user prototype and industrial case to present its executable functions. A system dynamics (SD) simulation model is adopted to validate the BLC-IRS framework, the simulation results indicated that our proposed BLC-IRS can be implemented as an effective a SC disruption response measure. Our developed BLC-IRS contributes an executable SCRes digital solution with synthetic technologies as a reactive SCRes measure, enabled users to mitigate the firm and partial network level disruption in an agile and safe manner.

## 1 Introduction

Supply chains (SCs) are becoming more intricate with rising uncertainties and risks (Manners-Bell, 2017) due to globalization. Designed in the period of globalization and lean management, modern SCs are compelled to adapt to revolutionary trends such as the global pandemics (i.e., COVID-19), reginal conflicts and the technology revolution (i.e., Industry 4.0 and 5.0).

Corresponding author email: yahu@tcd.ie

In light of these new changes, explorations on deploying advanced digital technologies to enhance SCRes are widely conducted in academia and industry (Ivanov 2021).

SCRes refers to the abilities of a supply chain to endure, adapt, and bounce back from unforeseen disruption events, return to or even surpass its initial performance. Businesses may even gain competitive advantages if they recover faster than their competitors (Spieske and Birkel 2021). According to Chowdhury and Quaddus (2017), SCRes is a multifaceted, hierarchical structure with three main dimensions: supply chain design quality, reactive, and proactive capabilities. To achieve the resilience goals completely, SC networks need to be built to adapt (Chowdhury and Quaddus 2017), sustain disruptions (low vulnerability), and recover from disruptions rapidly and cheaply (high recoverability) (Hosseini et al. 2019).

SCRes can be achieved proactively (Tang, 2006b) or reactively (Cheng and Lu 2017) by either creating redundancy or increasing flexibility (Sheffi and Rice Jr, 2005, Simchi ‐ Levi et al., 2018) through (1) readiness, (2) response, (3) recovery (Pyke and Tang, 2010), and (4) redesign stages (Hohenstein et al. 2015; Blackhurst et al. 2005) with internal or external collaboration (Ali et al. 2021). Proactive strategy is method that uses flexibility and redundancy components to create readiness before disruption.These elements usually include backup suppliers, alternate transportation routes, and capacity/inventory buffers (Ivanov et al., 2017). A reactive approach, on the other hand, is a tactic used in reaction to the SC disruption. In order to recover and expand, the reactive strategy incorporates features of redundancy, agility, and flexibility, including regionalization, capacity expansion, product and process change, and multi-sourcing (Ivanov et al., 2017; Hu and Ghadimi, 2023).

Taking proactive measures to prepare in the initial phase is the fundamental method for developing a resilient supply chain (Pyke and Tang, 2010). However, the unknown unknowns (Ivanov, 2024b), black swan events such as pandemics, severe natural disasters, terrorist strikes, or wars, are usually beyond the normal scope of expectations (Aven 2015). The 'tail-risk' scenario (O'Brien and MacAskill 2022) was not adequately anticipated by management procedures, as the effects of these occurrences are severe and totally unexpected (Spieske and Birkel 2021). For example, the proactive SCRes strategies solely in the readiness stage are inappropriate in face of the rapidity of the COVID-19 disruption (Ali et al. 2021). In this case, proposing reactive resilience strategies becomes essential.

As the SC performance declines rapidly after disruption in a concise time (Sheffi and Rice Jr 2005), illustrated in Figure 1, reducing the response time (Hohenstein et al., 2015) is one of the feasible SCRes measures after disruption. Nevertheless, attempts to develop SCRes strategies from the reactive aspect remain limited (Ivanov et al. 2017; Hu and Ghadimi 2023). Explorations to develop agile SCRes tools to respond to the unexpected SC disruption among the three reactive phases are insufficient. This is the first

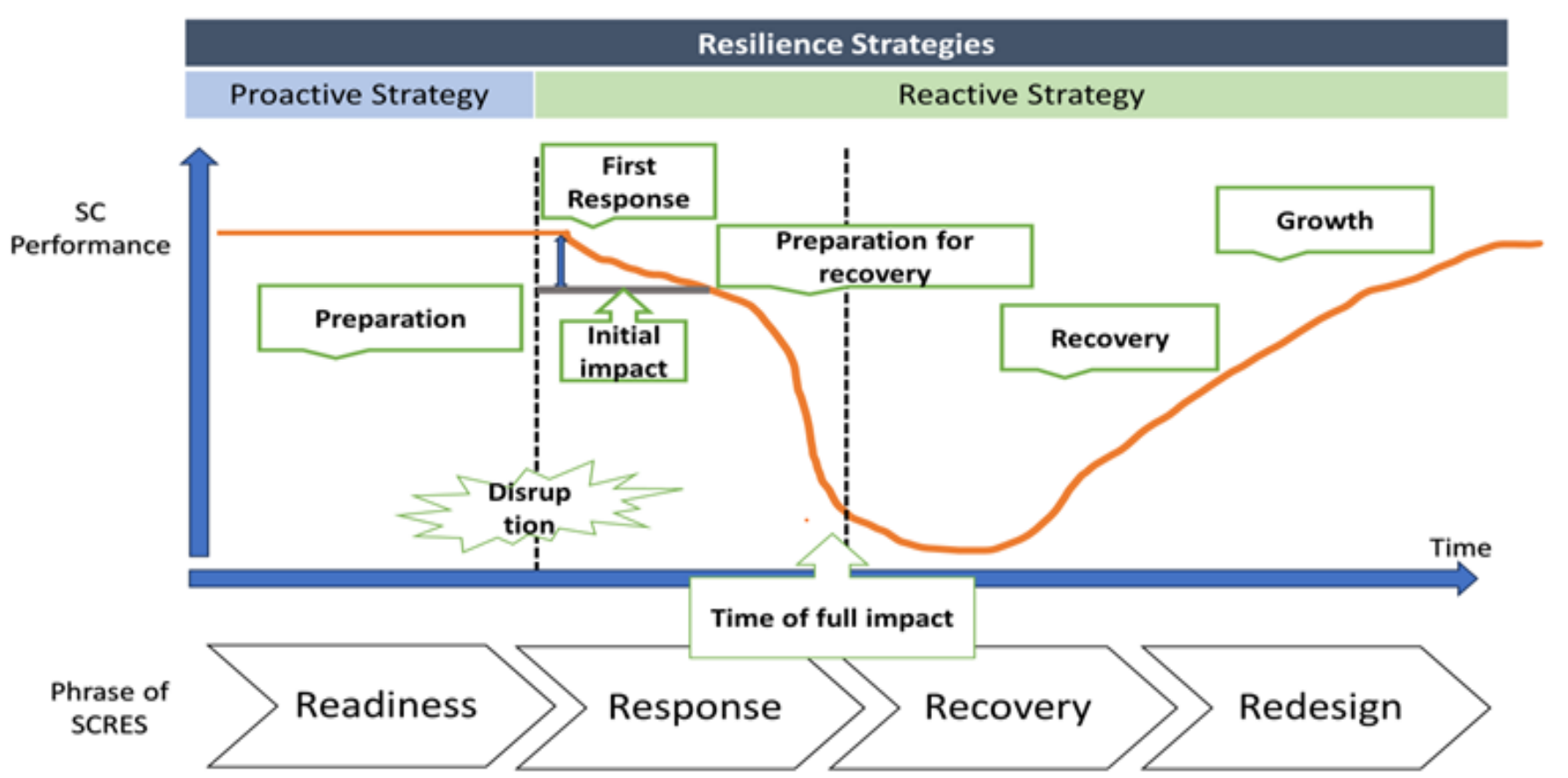

Figure 1: SC performance in different stages (Figure made by authors).

gap this research will address.

In the disruption response phase, businesses will use internal resources, such as internal inventory and capacity buffers, to fulfill the shortage during the response phase (Ivanov et al., 2017), and the SC performance declines somewhat gradually (Sheffi and Rice Jr 2005) due to the initial response (see Figure 1). When the internal resources were exhausted, SC performance declined rapidly, this is because the recovery preparatory tasks, such as increasing capacity or adjusting processes/products, take a relatively long time (Ivanov et al., 2017), the internal buffer resource has been exhausted and new supplemental resources owned by the same disrupted entity have not yet become available, the time interval makes the SC performance curve drops dramatically between the first internal response stage and the recovery stage. Under these circumstances, a rapid filtering, recommending, and visualizing tool to identify and demonstrate available external resources within the current SC network is an ideal solution to address this problem. The recommendation system (RS) can be a feasible solution as a SC disruption response measure.

A recommender system (RS) is a digital tool that identifies and selects the most suitable items or services (Chiu et al., 2021) for users, using current available information about users and items to estimate the utility of each associated item/service (Dadouchi and Agard, 2021). This is usually achieved by selecting relevant data from a large database (Yassine et al. 2021). The RS directly supports users in decision-making and addressing their information needs (Rana and Jain 2015) with attention to accuracy, context, novelty, real-time, and diversity dynamics.

Besides the basic information filtering and information matching function inherited from the RS, an intelligent recommender system (IRS) that embedded artificial intelligence (AI) algorithms also obtains a set of intelligent behavior and capabilities, such as information (knowledge) representation (clustering), learning, optimization, and reasoning mechanisms (Borràs et al. 2014; Aguilar et al. 2017). These capabilities enable the IRS to exploit, update, and infer knowledge (Aguilar et al., 2017). The IRS can be implemented in supply chain management (Pereira et al., 2022), improving deal speed (Sinha and Dhanalakshmi, 2019) and capturing dynamics (You et al., 2019).

SCRes can be achieved by increasing flexibility (Tang and Tomlin, 2008), and the high supply chain flexibility can come from an agile supply chain information system (Gupta et al., 2019). The principal mechanism for harnessing an IRS to develop SCRes is that this system is capable of tackling the challenges related to the rapidly expanded volume of information (Dadouchi and Agard, 2021) at high speed or in a real-time manner.

The IRS can help SC participants make well-informed decisions to use the available resources within the current supply network without acquiring extra resources (Dadouchi and Agard 2021) outside in a short timeframe. In this way, the disrupted entity can achieve resilience in an earlier stage by reducing the time interval between the SC disruption response and recovery phases by fast detection and resource utilization within the SC network with the assistance of the IRS.

However, explorations on RS in the supply chain management domain are scarce. Discussions on utilizing the IRS as an SCRes measure after SC disruption risk are still in their infancy. This is the second research gap this research will address.

The commonly adopted structure of information-sharing systems, such as IRS is centralized to save cost, streamline IT management and operations (Akinade, 2024). However, the centralized structure is more vulnerable to exogenous risks such as cyberattacks (Wan et al., 2020), the vulnerability of a centralized IRS will undermine the capability of our proposed IRS as a virtual SCRes tool. The BLC-based supply chain information systems have proven to have a higher resilience level as a real-world information system application (Cui and Idota, 2018, Enayati et al., 2024). Blockchain is a decentralized, distributed ledger and database that is immutable, transparent and traceable (Li et al., 2022). They have been adopted in several industrial supply chains for accessing correct and verified information, tracking the record changes transparently, implementing collaborative information governance, and protecting data privacy (Wan et al., 2020). Accordingly, blockchain can be an enabler for fostering SCRes (Hu and Ghadimi, 2022). Nevertheless, how blockchain can improve information sharing for complex multi-level SCs remains implicit (Guggenberger et al., 2020, Enayati et al., 2024). Studies for the synthetic utilization of BLC with RS to enhance SCRes remain unexplored. This is the third gap this study will address.

The three research gaps discussed above can be synthesized and summarized as: The development on blockchain based resilient intelligent recommender system for SC disruption response is absent. To tackle the research question, 'How to develop a BLC-based resilient IRS for SC disruption response?', the present research first proposes an IRS baseline as a new attempt at an SCRes measure in the reactive aspect. To ensure the reliability of the proposed IRS as a well-functioned digital SCRes measure and consider the practical needs for data quality, safety, and privacy for generating valid recommendation results, we integrated the BLC technology with the IRS baseline, developed a BLC-based IRS to enable the full potential of the IRS as an agile, efficient, reliable and safe SCRes enhancement tool.

To our knowledge, most discussions on digital technology applications for enhancing SCRes are limited to the conceptual level (Ivanov, 2024a), with stand-alone utilization (Bayramova et al., 2021) in the proactive aspect (Hu and Ghadimi, 2024). The adoption of the IRS-based framework as a response measure after the SC disruption is limited, the integration of BLC and the IRS to build a resilient and safe digital SCRes decision support system is yet to come. The contributions of this research are twofold. This research offers a new application domain for the IRS and BLC techniques in addition to contributing to the SCRes body of knowledge with a focus on reactive SCRes strategies that have been neglected by previous studies.

The remainder of this paper is structured as follows: Section 2 presents a literature review on related works. Section 3 describes the IRS baseline and the complete BLC-IRS framework architecture, explains the system implementation steps. An implementation prototype of a smart contract is shown to illustrate the data exchange mechanism of the BLC-IRS. To demonstrate the user functions of the proposed BLC-IRS, an industrial use case was presented. A System Dynamics (SD) simulation is adopted in Section 4 to validate the BLC-IRS framework as a novel information-sharing mechanism with SC resilience development capability. The deployment scope of the BLC-IRS is presented. In Section 5, theoretical contributions, management insights, and discussion are provided. The conclusion and future directions for research are highlighted in Section 6.

## 2 Literature Review

To identify the research gaps in the current existing SCRes enhancement digital tools development literature, we searched the phrases 'supply chain resilience' AND 'digital technology' OR 'digital tools' in the article's title, abstract, and keywords in Scopus Database and Google Scholar since 2019. We added the key word search 'blockchain' AND 'supply chain resilience', 'artificial intelligence' AND 'supply chain resilience' ,'recommender system' AND 'supply chain resilience' to refine the clarity, the most related application-oriented research works with direct digital solutions were divided into four categorizations, (1) action aspects of the studied SCRes strategies to understand if the developed tools functioned in a proactive or reactive aspect, (2) applied techniques to enhance SCRes, (3) application domain and (4) phase of SCRes. The comparison between related research works is illustrated in Table 1.

Table 1: Techniques applied to enhance SCRes in various phases.

| Reference | Action Aspect | Techniques | Application Domain | Phase of SCRes |
|---|---|---|---|---|
| (Abadi et al., 2024) | Proative+ Reactive | Blockchain + Machine Learning | Resilient decision support system development for dynamic pricing | Readiness +Response |
| (Aboutorab et al., 2024) | Proactive | Natural Language Processing +Reinforcement Learning | Disruption identification | Readiness |
| (Al-Haidous et al., 2022) | Proactive | Mathematical Model+ Simulation | Resilient Network Design | Readiness |
| (Ashraf et al., 2024) | Proactive | Deep Learning | Disruption Prediction | Readiness |
| (Badakhshan and Ivanov, 2025) | Proactive+ Reactive | Digital Twin+Blockchain+Machine learning | Disruption impact evaluation+Disruption response | Readiness +Response |
| (Baležentis et al., 2021) | Proactive | Simulation | Resilient Strategy Assessment | Readiness |
| (Belhadi et al. 2022) | Proactive | Fuzzy Wavelet Neural Network (FWNN) | Resilient Strategy Assessment | Readiness |
| (Caputo et al. 2023) | Proactive | Mathematical Model | Resilience Level Assessment | Readiness |
| (Cavalcante et al. 2019) | Proactive | Machine learning +Simulation | Resilient supplier selection | Readiness |
| (Chen and Chen 2023) | Proactive | Mathematical Model | Resilient Network Design | Readiness |
| (Cuong et al. 2023) | Proactive | Deep Learning | Disruption Prediction | Readiness |
| (Enayati et al., 2024) | Proactive | Blockchian+Simulation | Resilient Information System Development | Readiness |
| (Habibi et al., 2023) | Proactive | Simulation | Resilient Strategy Assessment | Readiness |
| (Habibi et al., 2025a) | Proactive | Simulation | Resilient Strategy Assessment | Readiness |
| (Habibi et al., 2025b) | Proactive | Simulation | Resilient Strategy Assessment | Readiness |
| (Hosseini and Ivanov 2022) | Proactive | Mathematical Model+ Simulation | Disruption impact evaluation | Readiness |
| (Hosseini et al. 2019) | Proactive | Mathematical Model | Resilient supplier selection and optimal order allocation | Readiness |
| (Ivanov 2023) | Proactive+ Reactive | Digital Twin, AI in general | Resilience Level Assessment | Readiness +Response |
| (Kang and Bhawna, 2025) | Proactive | Machine Learning | Resilient supplier selection | Readiness |
| (Li et al., 2025) | Proactive | Deep learning | Disruption prediction | Readiness |
| (Lin et al., 2025) | Proactive+ Reactive | Generative AI+Mathematical Model | Disruption prediction+ resilient strategy selection | Readiness +Response |
| (Long et al. 2023) | Proactive | Echo state network Model (ESN), Artificial Neural Network | Disruption Prediction | Readiness |
| (Lorenc and Kuźnar 2021) | Proactive | Artificial neural networks | Disruption prediction | Readiness |
| (Manzoor et al., 2025) | Proactive | Mathematical Model+Blockchian | Resilient supplier selection and optimal order allocation | Readiness |
| (Mohammed et al. 2021) | Proactive | Mathematical Model | Resilient supplier selection | Readiness |
| (Pachot et al., 2021b) | Proactive +Reactive | Natural Language Processing +Recommendation System | Resilient supplier recommendation | Readiness +Response |
| (Parhi et al., 2024) | Proactive | Mathematical Model+Simulation | Resilient Strategy Assessment | Readiness |

| (Sani Mohammed et al. 2023) | Proactive | Simulation | Resilient Strategy assessment | Readiness |
|---|---|---|---|---|
| (Silva et al. 2023) | Proactive | Simulation | Resilient Network Design | Readiness |
| (Singh et al. 2019) | Proactive+ Reactive | Semantic technology and ontology knowledge base+ Mathematical Model | Resilient decision support system development to determine an optimal recovery activity | Readiness +Response |
| (Singh et al., 2025) | Proactive+ Reactive | Blockchain+Multi agent Deep reinforcement Learning | Dynamic demand forecasting | Readiness +Response |
| (Tsiamas and Rahimifard 2021) | Proactive | Simulation | Disruption impact Evaluation | Readiness |
| (Yang et al., 2024) | Proactive+ Reactive | Machine Learning | Post-disruption sales prediction | Readiness +Recovery |
| (Zhang et al., 2024) | Proactive+ Reactive | Simulation+Mathematical Model | Disruption Impact Evaluation+ Resilient Strategy Selection | Readiness +Response |
| Present study | Proactive+ Reactive | Intelligent Recommendation Systems+Blockchain | Resilient decision support system development for resilient resource recommendation | Readiness +Response |

From the action aspect perspective, research focusing on developing SCRes measures from a reactive aspect and at the response phase is limited, although the supply chain performance drops rapidly after disruption. There are 9 out of 34 reviewed papers proposed digital SCRes solutions considering both proactive and reactive aspects, in the readiness and response phases. Only 2 works proposed resilient decision support system development after SC disruption, one for dynamic pricing (Abadi et al., 2024), the other one for optimal recovery activity determination (Singh et al. 2019). Share the most similar concept with study from Singh et al. (2019), the present research illustrated our distinctiveness by a) leveraging the advanced BLC, IRS digital techniques compared to their ontology-based decision support system; b) considering both internal and external resilient resource recommendation, while the previously developed ontology-based decision support system suggested the new values for the parameters such as production capacity of units, transportation capacity only use firm's internal data. Our proposed solution obtains higher flexibility and scalability.

Upon the techniques' usage and application domain, mathematical models (Chen and Chen 2023; Caputo et al. 2023), simulation (Silva et al. 2023; Sani Mohammed et al. 2023) including digital twin (Ivanov 2023), and fuzzy logic (Belhadi et al. 2022) are widely depoyed for SCRes assessment (Caputo et al. 2023; Sani Mohammed et al. 2023; Belhadi et al. 2022; Habibi et al., 2025a), resilient supplier selection (Mohammed et al. 2021; Cavalcante et al. 2019), resilient SC network design (Chen and Chen, 2023; Silva et al. 2023) and disruption impact evaluation (Tsiamas and Rahimifard 2021; Hosseini and Ivanov 2022). AI techniques such deep learning (Cuong et al. 2023; Li et al., 2025) and artificial neural networks (Lorenc and Kuźnar 2021; Long et al. 2023) were adopted to forecast disruptions. Machine learning was also capable on post-disruption sales prediction (Yang et al., 2024), resilient supplier selection (Kang and Bhawna, 2025). Reinforcement learning combined with natural language processing (NLP) were used for disruption identification (Aboutorab et al., 2024). NLP techniques were also found to be integrated with recommender system for resilient supplier recommendation (Pachot et al., 2021b). Generative AI was integrated with mathematical models to accomplish disruption prediction and resilient strategy selection (Lin et al., 2025). In recent studies, the BLC was found to be implemented to develop a resilient supply chain information system (Enayati et al., 2024). BLC was also synthetic used with AI to develop a resilient decision support system (Abadi et al., 2024), secure the transaction data (Badakhshan and Ivanov, 2025) for dynamic demand forecasting (Singh et al., 2025). One research attempted to develop a resilient decision support system with semantic technology and ontology techniques (Singh et al., 2019).

From the techniques' adoption perspective, an RS can assist resource-intensive supply chain procedures (van Capelleveen et al. 2021) by increasing item/service explorations and lowering search costs for locating pertinent prospects. Capabilities such as information (knowledge) representation (clustering), learning, optimization, and reasoning mechanisms (Borràs et al. 2014; Aguilar et al. 2017) obtained by an IRS enable the feasibility of leveraging the IRS as an effective tool for mitigating SC disruption risk. However, investigations on the possibilities of such a development are still lacking, despite the IRS's agility capabilities enabling it to become a useful instrument for mitigating SC disruption risk. Only one paper in the reviewed literature discussed harness IRS to tackle resilient supplier recommendation challenges (Pachot et al., 2021b).

From the application domain perspective, it is worth noting that leveraging an IRS technique as a SCRes enhancement tool to form a resilient decision support system, particularly for agile responses, is often neglected. Furthermore, to enhance the reliability, stability, and data quality of an IRS, integrating BLC technology with the IRS presents a unique opportunity to develop a resilient decision support system framework. Such integration is rare in the reviewed literature. Our present study shared the similar multi-objective recommendation system (Pachot et al., 2021a) concept as the fundamental techniques with the study from Pachot et al. (2021b), to our knowledge, this study is the first attempt that integrate the BLC with IRS to establish a resilient decision support system for SC disruption response.

Based on the discussions above, the research gaps can be summarized:

1). Digital adoptions for strengthening SCRes have been predominantly pursued from a proactive aspect, reactive digital solutions after disruptions are insufficient.

2). The adoption of the IRS as a reactive SCRes measure is inadequate.

3). The deployment studies on the integration of BLC technology with an IRS to form a resilient decision support system for SC disruption response considering on both internal and external resilience resource suggestion remain unexplored (Hu and Ghadimi, 2022).

To address these identified gaps, the present study developed an BLC-IRS as a resilient SC disruption response system for both internal and external resource recommendation.

## 3 The BLC-IRS SC Disruption Response System Framework Design and Implementation

The overall research procedure (Peffers et al., 2007) and research methods for developing such a BLC-IRS information system as the SC disruption response measure are illustrated in Figure 2.

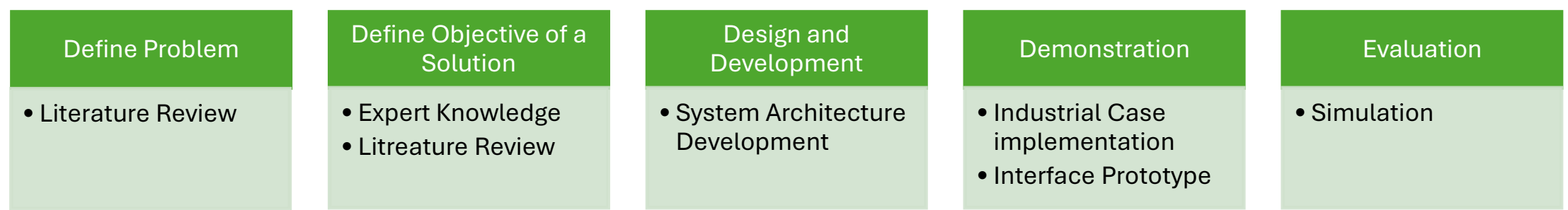

Figure 2: Research procedure and method overview.

The problems to be addressed and the objective of developing the BLC-IRS were defined in Section 2. In this Section, the IRS baseline framework was proposed to illustrate the core component and functions of the BLC-IRS, then the completed BLC-IRS system architecture was developed and demonstrated by industrial case and user interface prototypes. The BLC-IRS was evaluated by the simulation method in Section 4.

### 3.1 The Baseline Framework Overview and Description

The IRS baseline is designed to mitigate the SC disruption risk by improving response speed after disruption. The IRS was developed as an SCRes enhancement tool to enable a quick response to SC disruption, as seen in Figure 3. The function mechanism of the proposed IRS baseline is illustrated in this section. Following the function procedure of the baseline framework, the internal workflow of the proposed IRS is explained in Figure 4.

The following outlines the procedures of the developed IRS:

Step 1: Determine the available internal redundant resources and suggest them as an initial response to mitigate SC disruption.

Step 2: Determine and identify accessible external redundant resources to mitigate the SC disruption between the initial response and recovery phases after the internal resources are exhausted.

The sequence of recommendations is from step 1 to step 2. When the SC is disrupted, the IRS will first identify and recommend the available internal resources to the disrupted entity. And after the internal resources are exhausted, the IRS will go to detect accessible external redundant resources within the SC network to tackle the shortage problem faced by the disrupted entity.

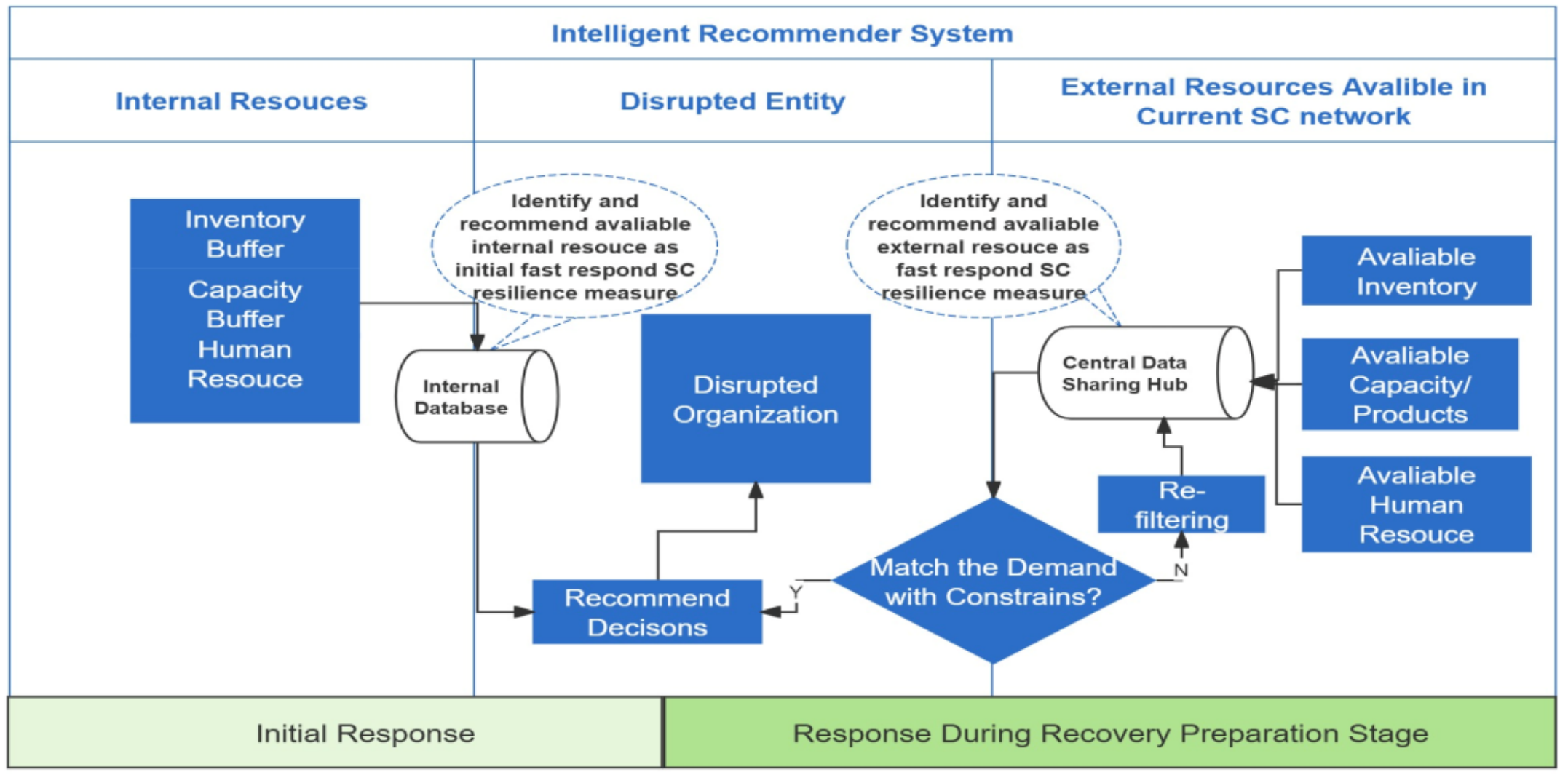

Figure 3: The proposed IRS Baseline Architecture.

Step 1 represents the internal resource recommendation function during the initial response phase. Information about available internal redundant resources, such as inventory/capacity buffers, human resources, will be filtered out from the disrupted entity's internal database during the initial response period for disruption mitigation. When internal resources are exhausted after the first response, the disrupted entity will turn to acquire accessible external resources within its supply network. This can be a solution to address the present shortage with higher efficiency, as building new supplements on its own would take a relatively longer time.

Following the first internal response action, Step 2 shows how the IRS recommends external resources during the response phase. In order to find available external resources inside the same supply network, the system filters and matches shared data from the central database, such as current available inventory utilization rate, excess production capacity percentages, workforce availability and lead time. These quantified resource attributes allow the IRS to recommend feasible external resource owners to mitigate the shortage in the response phase before the new internal capacity was established.

The internal recommendation is straightforward, while the external recommendation results will be generated by combining the practical conditions, such as the constraints on lead time, emergency cost, transportation channel, or other requirements will be made by the user (disrupted entity) in an actual operation environment. Since users' preferences for choosing proper external resource owners will vary depending on different disruption scenarios and specific needs. In some urgent case, the lead time criteria will be considered as the most critical factor in mitigating a sudden disruption. In other scenarios, the emergency cost can be recognized as the most important criterion before generating recommendation results, as the shortage problem at that time is not entirely urgent. The IRS will not stop filtering and searching until all the requirements are fulfilled. And after all the actual needs are met, the recommendations will be finalized, and the results will be generated. This IRS is distinctive for considering the practical limitations of the existing SC network, which allows SC practitioners to make better decisions based on reality, both internal and external. The proposed IRS is a useful SCRes measure because of its practical functionality. Moreover, it is possible to execute real-time and dynamic recommendations to better capture the complexity and dynamic characteristics of the SC.

The internal workings of the proposed IRS were described in Figure 4. Both static and dynamic real-time data can be used with this IRS. The inner workflow is constructed with three components: (i) data preparation and processing, (ii) recommendation algorithms selection, and (iii) recommendation services promotion. In the data processing stage, the data will be input, cleaned, and used to create a basic profile of external resources. The external resource owner's profile will be utilized to match with the current user profile after it has been generated, since this is the primary working mechanism of all types of RSs. The second step involves determining algorithms to finish the profile matching between the users and the external resource. Algorithms will be determined depending on the practical problem and data features. Classical RS algorithms such content-based, collaborative, and knowledge-based recommendation algorithms can be found in the IRS. AI-based algorithms such as machine learning, deep learning, and artificial neural network (ANN) are also embedded in the IRS algorithm engine.

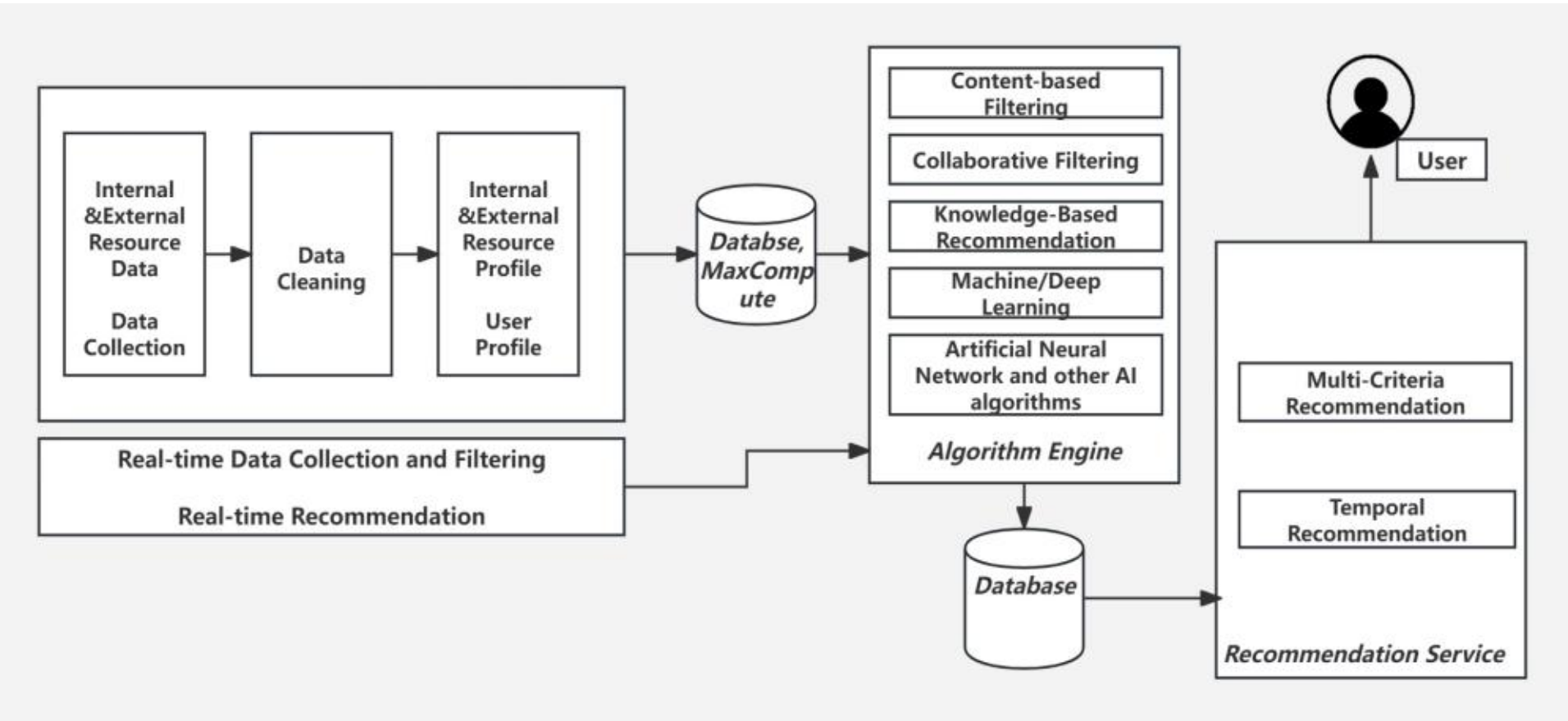

Figure 4: Inner workflow of the IRS.

Recommendation service execution is the last step of this workflow. The multi-criteria and temporal recommendations will be used as the main recommendation, considering the real-time requirements and practical limitations on the users' preferences.

### 3.2 Centralized Information System Versus Decentralized Information System

The information system architecture illustrated in Figure 3 is a centralized information-sharing scheme between the disrupted organization and the external resource owners.

Normally, a centralized SC information system offers cost savings, budget control, and streamlined IT management and operations. With a centralized information-sharing scheme, participants can implement standardized security protocols across the SC (Akinade, 2024). However, a centralized information-sharing structure can also pose challenges in communication and reduce responsiveness to localized demands (Akinade, 2024). The hierarchical relationship between upstream and downstream SC participants can result in the bullwhip effect (Xue et al., 2021), which typically leads to exaggerated orders for upstream suppliers (Lee et al., 1997, Liu et al., 2024). In our case, the downstream customer with emergency demand is the user, the disrupted organization, and the upstream suppliers are the external resource owners. In the scenario of emergency order placement after the SC disruption, the exact and specific order information flow is essential. Any increased production compliance caused by the order exaggeration resulting from the centralized SC information sharing scheme will lead to longer processing times for order preparation among external suppliers, thereby decreasing reaction efficiency. This is contrary to the need for rapid response to address shortage problems after SC disruption. In this case, the centralized structure will undermine the capability of the proposed IRS as a fast response SCRes measures. The degree of information sharing will be significantly reduced by the centralized information transmission and further reduce the supply chain flexibility, seriously jeopardizing the cooperation enthusiasm of SC participants, and negatively impacting synergy (Xue et al., 2021). Moreover, the centralized database system is vulnerable to information leakage and cyberattacks; it also cannot trace alterations to sensitive data stored (Wan et al., 2020). These disadvantages of the centralized information-sharing scheme render our proposed IRS insufficiently resilient to perform as a stable and reliable SC resilience tool. In this case, a decentralized information-sharing scheme can be considered an alternative.

The characteristics of the decentralized systems such as fault-tolerant, location-independent, and extensible (Xue et al., 2025) are almost exact opposite to centralized systems. Under a decentralized information-sharing scheme, the SC participants can directly use and better control the data that is most relevant to them, therefore substantially lowering communication costs (Rockart and Leventer, 1976, Xue et al., 2025).

Moreover, the decentralized information-sharing scheme obtains higher resilience in the face of exogenous risk or threat due to the distributed communication structure illustrated in Figure 5.

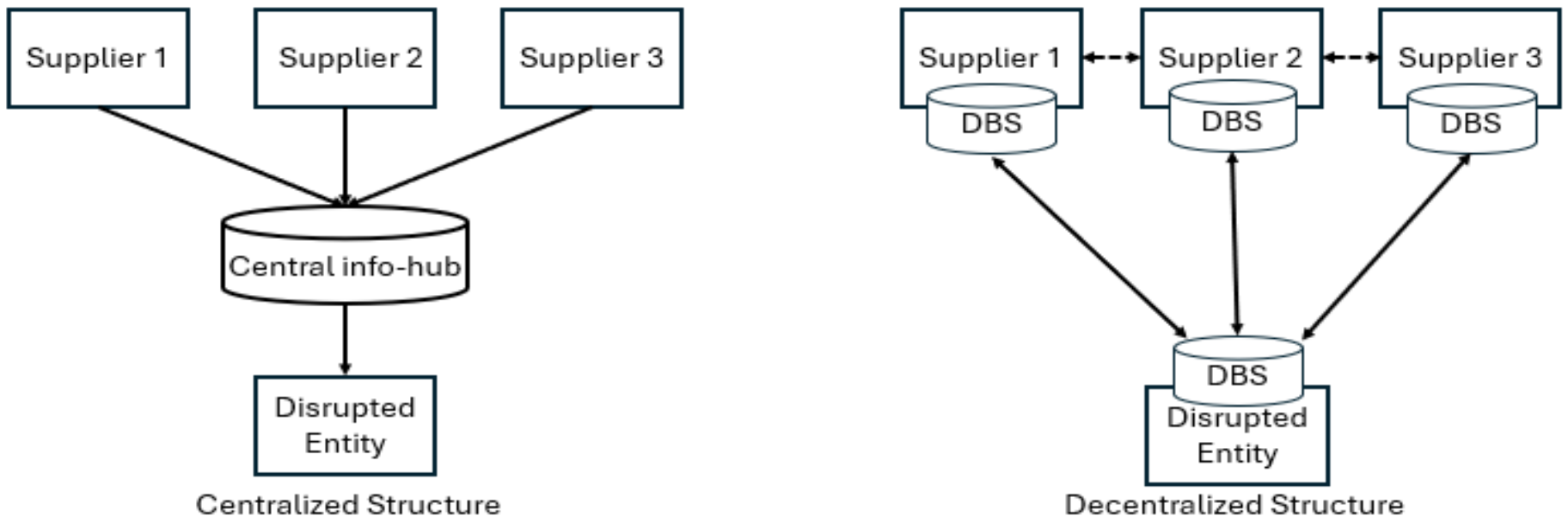

Figure 5: Centralized and Decentralized Information-sharing structure.

### 3.3 The potential Drawbacks of the Proposed IRS baseline

The centralized communication structure is vulnerable to disruption risk. In our case, if the central information hub was disrupted (See Figure. 5), the entire IRS would malfunction as there would be no input data for generating recommendation results. On the contrary, the decentralized design of the communication scheme between the disrupted entity and external resources can mitigate the external risk that disrupts data communication, as alternative communication channels are usually available. Due to the peer-to-peer communication mechanism within the decentralized structure, a commonly recognized truth can be built, whereby the information cannot be manipulated by an individual (Tiwari et al., 2023), leading to a higher level of trust and willingness to collaborate among SC participants. Higher motivation in collaboration is an important antecedent of SCRes (Scholten and Schilder, 2015). Moreover, decentralized systems have been proven to be more robust than centralized systems against external cyber-attacks (Chen et al., 2021), decentralized management enables the network structure better realize complementary advantages , reduces risks of information system and the SC (Xue et al., 2021).

The proposed IRS baseline is designed for fast response after SC disruption based on efficient, agile information exchange and matching capability. In a very short timeframe, more frequent interactions with surging data volume among different entities in the supply chain network make higher requirements on the supply chain information system (Cui and Idota, 2018). A resilient information-sharing scheme is the fundamental to a well-functioning and long-lasting IRS after disruption that can ultimately enhance SC resilience at the physical level. If the information exchange scheme between the disrupted entity and external suppliers is not stable or can be disrupted easily, the information matching function cannot be achieved. Therefore, users cannot get the recommendation results from the most suitable external resources. Moreover, even in the case of a stable information-sharing scheme, the recommendation results will be determined by the quality and authenticity of the data provided by external entities. In the context of data-driven IRS, the importance of information system security and reliability cannot be overstated (Putro et al., 2023). A reliable information-sharing scheme between disrupted entities and external entities with trustworthy data is essential for achieving the full resilience capability of the proposed IRS. Moreover, besides the static data exchange, the real-time data exchange might be another user requirement depending on the actual needs after SC disruption. From the perspective of a decentralized information system that enables commitment to information quality, security, and real-time operation, the BLC-based IRS is an ideal solution to developing a resilient IRS at the top of the proposed baseline presented in Subsection 3.1.

### 3.4 The Feasibility of BLC Integration

Blockchain is a distributed shared ledger and database that is decentralized, immutable, traceable (Li et al., 2022). The BLC can change and reestablish the connections among all supply chain system participants based on these functions (Xue et al., 2021). This has been adopted in several industrial supply chains, such as medical and health, construction, and textile industries, for accessing correct and verified information, tracking the record changes transparently, implementing collaborative information governance, and protecting data privacy (Wan et al., 2020). Besides the discussions on the conceptual level, discussions on the BLC integration with the SC information system at the implementation level were also raised. By using BLC, Cui and Idota (2018) reconfigured the vegetable SC information system, and found that SC will become more transparent, enabling broader collaboration and simplifying the process, therefore increasing the resilience level. Enayati et al. (2024) illustrated that the integration of BLC technology supply chain information systems enhance the reliability, efficiency, and security of information exchange among SC participants, therefore reinforcing the SCRes by presenting a regional case. In the industrial application aspect, the IBM Food Trust project leverages BLC to enhance transparency and reduce SC risk (Sodhi and Tang, 2019). Users can trace products across the SC in a real-time manner, and SC participants can deploy smart contracts within private channels involving two or more designated network members. Through this system, decision-making processes can be automated (Kawaguchi, 2019). BLC is capable of enhancing the resilience of the SC information system in the cyber environment through safeguarding data privacy, securing data storage (Ray et al., 2024) and improving visibility by sharing cybersecurity-related information among supply chain participants (Hämäläinen and Nadesan, 2022).

Discussions on developing SCRes by integrating the BLC technologies are mostly limited to the conceptual level, with qualitative analysis (Ivanov, 2024a). The applications of BLC adoption SC in real-world contexts remain insufficient (Bayramova et al., 2021). The decentralized nature obtained by the BLC can help eliminate the exogenous threat to the SC information network, proving that the implementation of BLC can be an antecedent of developing a resilient supply chain information system (SCIS) (Bayramova et al., 2021). However, studies on the development and implementation of BLC-based SCIS are insufficient. Moreover, most previous studies focused on stand-alone blockchain applications (Bayramova et al., 2021), the utilization of the BLC and other advanced digital tools to solve the SCRes problem is insufficient (Hu and Ghadimi, 2022). The combination of the BLC (Parlak, 2025) and the IRS that worked as a resilient decision-support system for SC disruption response is still in its infancy. This research addressed this gap by proposing a BLC-IRS framework for SC disruption response.

### 3.5 BLC-IRS Framework Design

In this subsection, we describe the structure and function of the proposed BLC-IRS, present the architecture of the BLC-IRS framework, and implement a smart contract to illustrate the information exchange mechanism in the proposed BLC-IRS.

#### 3.5.1 The Decentralized System Structure

The decentralized nature of BLC eliminates the dependence on one particular information centre that governs sensitive information therefore enhancing trustworthy and collaborative partnerships (Wan et al., 2020) among different SC participants, while a higher collaboration level can increase SCRes (Scholten and Schilder, 2015) and decrease transaction cost (Wan et al., 2020). The decentralized structure of BLC-based information systems can mitigate risks associated with instability of the central information hub, such as cyberattacks and compromised privacy. The BLC-based information-sharing scheme is also capable of encrypting touchy information, which can protect the database (Hashim, 2018). The reliability and security characteristics of the BLC-based information-sharing scheme are the essential requirements of a functional IRS deployed as an SCRes enhancement tool. The implementation of blockchain-driven technology solidifies trustworthy and safe transactions by creating an immutable and transparent log, mitigating threats, and enhancing traceability (Tiwari et al., 2024).

### 3.5.2 The Smart Contract and Traceability Application

The smart contract is one of the representative blockchain applications in the SCRes domain (Min, 2019, Hu and Ghadimi, 2022, Beck et al., 2023). It can execute the contractual agreement automatically with enforcement as it can self-verify and self-execute agreements. This functionality streamlines the contract lifecycle to enhance compliance and mitigate risk. For the disrupted entity, the emergency procurement contracts (Omar et al., 2021) can be executed automatically according to the recommendation results. For the external suppliers, the billing transaction (Min, 2019) can be finished automatically in a secure, agile manner (Beck et al., 2023) after the replenishment is delivered successfully to the disrupted entity without human intervention. This can help to build trust (Dubey et al., 2017) among SC participants and to control and accelerate transaction executions, therefore improving collaboration and agility, ultimately increasing the overall SCRes levels (Scholten and Schilder, 2015, Aslam et al., 2020). By creating mutual benefits for the disrupted entity and external resource owners,  partners on both the demand and supply side can get information and perform transactions within a secure and trustworthy environment governed by the properly developed smart contracts (Grida and Mostafa, 2023).

The traceability function (Tang and Zimmerman, 2013) of a BLC-based information system can also be a driver of mitigating the overall supply chain risk (Hu and Ghadimi, 2022) from a long-term perspective, especially for the disrupted entity in our case.  An emergency order placement following a disruption is typically a short-term or one-time operation. The traceability feature (He et al., 2025) of the BLC-based information-sharing system can help the core customer, the disrupted entity, trace back the external supplementary information that was not of adequate quality or safety after a short-term or one-off but critical transaction. In this case, the unqualified external supplier can be identified, re-audited, re-rated (Liang et al., 2024), or excluded from the cooperation network, and the collaboration relationship network between the disrupted entity and external suppliers will be reconstructed. As the unreliable external suppliers were deprioritized or excluded from the cooperation list, from the long-term perspective, the disrupted entity will encounter less risk from the external cooperation thus building a more sustainable, reliable SC network.

### 3.5.3. BLC-IRS Framework Architecture and Smart Contract Implementation

The system architecture proposed in Section 2 has been modified and illustrated in Figure 6 based on the function descriptions in Subsections 3.5.1 and 3.5.2.

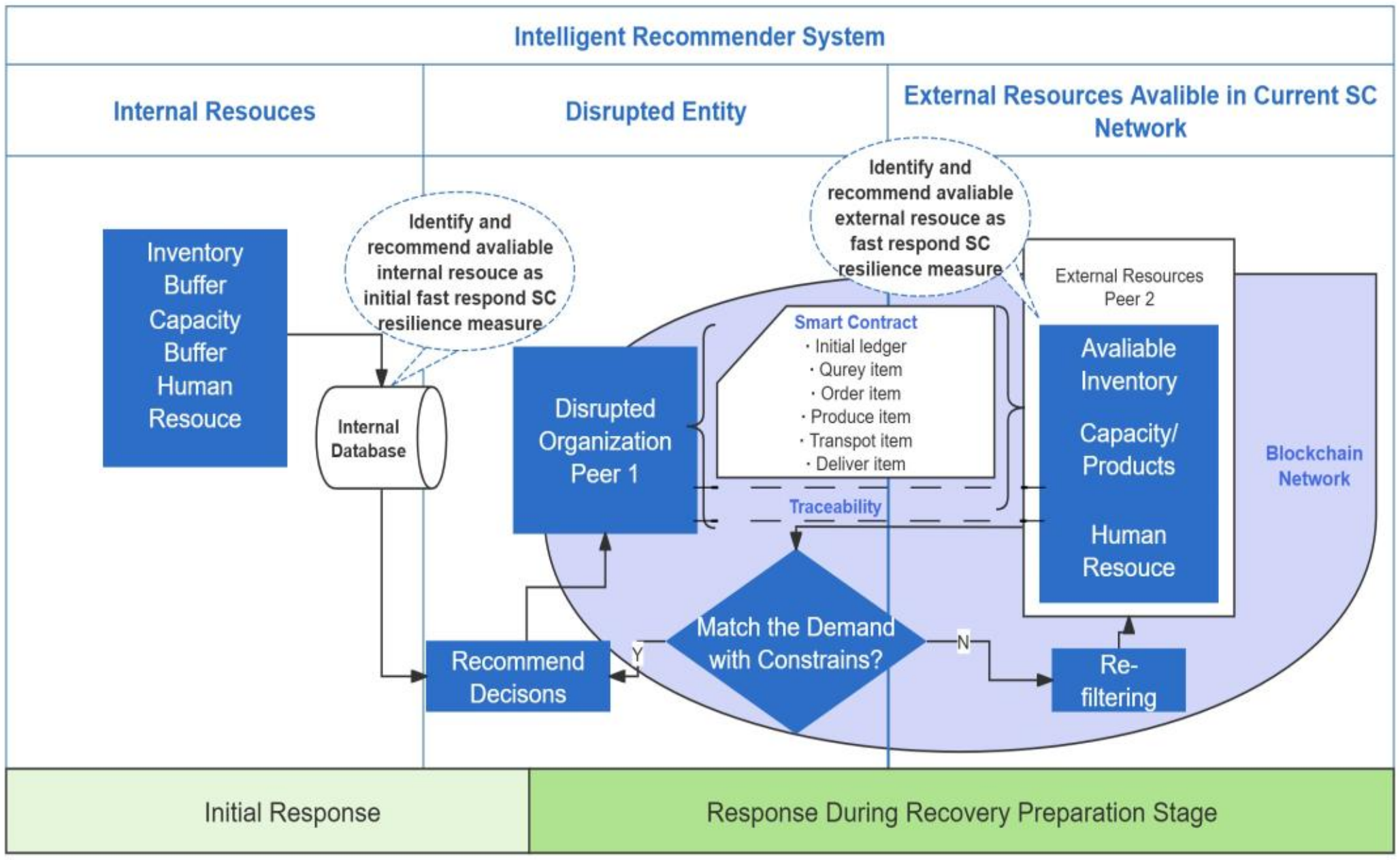

Figure 6: BLC-based IRS architecture (proposed by the authors).

Based on the architecture in Figure 6, this research designs a Produce Item Contract (PIC) as an example to realize the smart transaction and product traceability function. The PIC contract is deployed by the external resource owner, which stores information for their available resources in the same network. The external resource owners input their product information in the contract, including product type, SKU, supplier name, location, lead time, production volume, and price (see Table 2). The dataset used to register external resource information from Kaggle at https://www.kaggle.com/datasets/harshsingh2209/supply-chain-analysis.

Table 2: Example of external resource information list in PIC contract.

| Product Type | SKU | Supplier Name | Location | Leadtime | Production Volume | Price |
| --- | --- | --- | --- | --- | --- | --- |
| Haircare | SKU0 | Supplier 3 | A | 12 | 917 | 33 |
| Skincare | SKU1 | Supplier 3 | B | 24 | 937 | 30 |
| Cosmetics | SKU2 | Supplier 1 | C | 5 | 414 | 35 |

According to Wang et al. (2019), the PIC contract can be viewed as internal data exchange platform between the user and the external resource owners. The external resource information can be registered on the PIC contract platform. The PIC contract preserves the mapping link between the product category, SKU, and the Order Item contract address. The external resource owner registers their products, allowing other nodes, in this case, the disrupted entity, to capture the registered resource information through the BLC network. This information is then used to generate recommendation results based on the information

Production Info Contract Prototype

Add Product Info

Haircare
SKU0
Supplier 3
A
12
917
33
Add Product

Get Product Info

Enter Product ID
Get Product Info

Figure 7: Resource Owner interface page.

Get Product Info

1
Get Product Info
**Product ID:** 1
**Type:** Haircare
**SKU:** SKU0
**Supplier:** Supplier 3
**Location:** A
**Lead Time:** 12 days
**Volume:** 917 units
**Price per unit:** 33 EUR
**Created At:** 3/21/2025, 11:45:53 PM

Figure 8: Disrupted entity interface page.

mapping process, which mitigates the shortage problem caused by the SC disruption. Figures 7 and 8 illustrate the prototype mechanism and user interface of the PIC contract. The smart contract was developed at the browser-based open platform Remix. Remix is an Ethereum Solidity language development tool for smart contract development and supports test functions. The user interface prototype is created by first developing the web page using HTML and CSS code and then connecting the HTML code with front-end JavaScript code to create the front-end web page's interaction logic. By clicking on the button, users can engage with contracts, deploying new contracts and using the contract's mechanisms to read or write blockchain data.

A resource registration page for the externally available product is presented in Figure 7. This page contains a list of all registered items and raw material details in the system. By entering the necessary resource details, the external resource owner can register their products on the blockchain, establish information source for their goods, and create unchangeable records. At the demand side for the external resource information, the disrupted entity can acquire the external resource information immediately after the registration at the supply end, illustrated in Figure 8. The BLC components can help resist information poaching and manipulation (Bossler et al., 2024). Once the supply information is uploaded, it cannot be revised without consensus. This commitment to data authenticity is critical and fundamental for generating valid recommendation results. If external resource information can be inputted in real-time by the external resource owners, the disrupted entity can capture the real-time data and generate recommendation results in real-time at the demand side. Similarly, the data provided by the supply end cannot be deleted without consensus for the disrupted entity. This facilitates the traceability function to build a sustainable and low risk supply network in the long term.

This BLC-based information-sharing can commit to the security and reliability of the IRS baseline system due to the decentralized structure. With the smart contract, the BLC-based IRS can promise the information passed by external suppliers with authenticity, privacy, security, and real-time visibility (Beck et al., 2023). Therefore, the recommendation results were generated without any delay in system response time. The embedded traceability function of the BLC-based IRS can help the disrupted entity re-access external suppliers and build a more sustainable and low-risk SC in the long term.

The fast response speed at the information level by implementing the IRS is the key mechanism to achieving SCRes at the physical level. The integration of the BLC technology will not affect the response efficiency of the proposed IRS as the BLC-based information system is widely recognized to be agile or real-time for automation (Beck et al., 2023). Synthetic implementation with the IRS and the BLC technology will make the IRS itself more stable and resilient as a digital SCRes measure and guarantee that the SCRes of the physical product level can be achieved via a fast, secure response decision support system at the information level. The decision process of the BLC-IRS was illustrated in Figure 9.

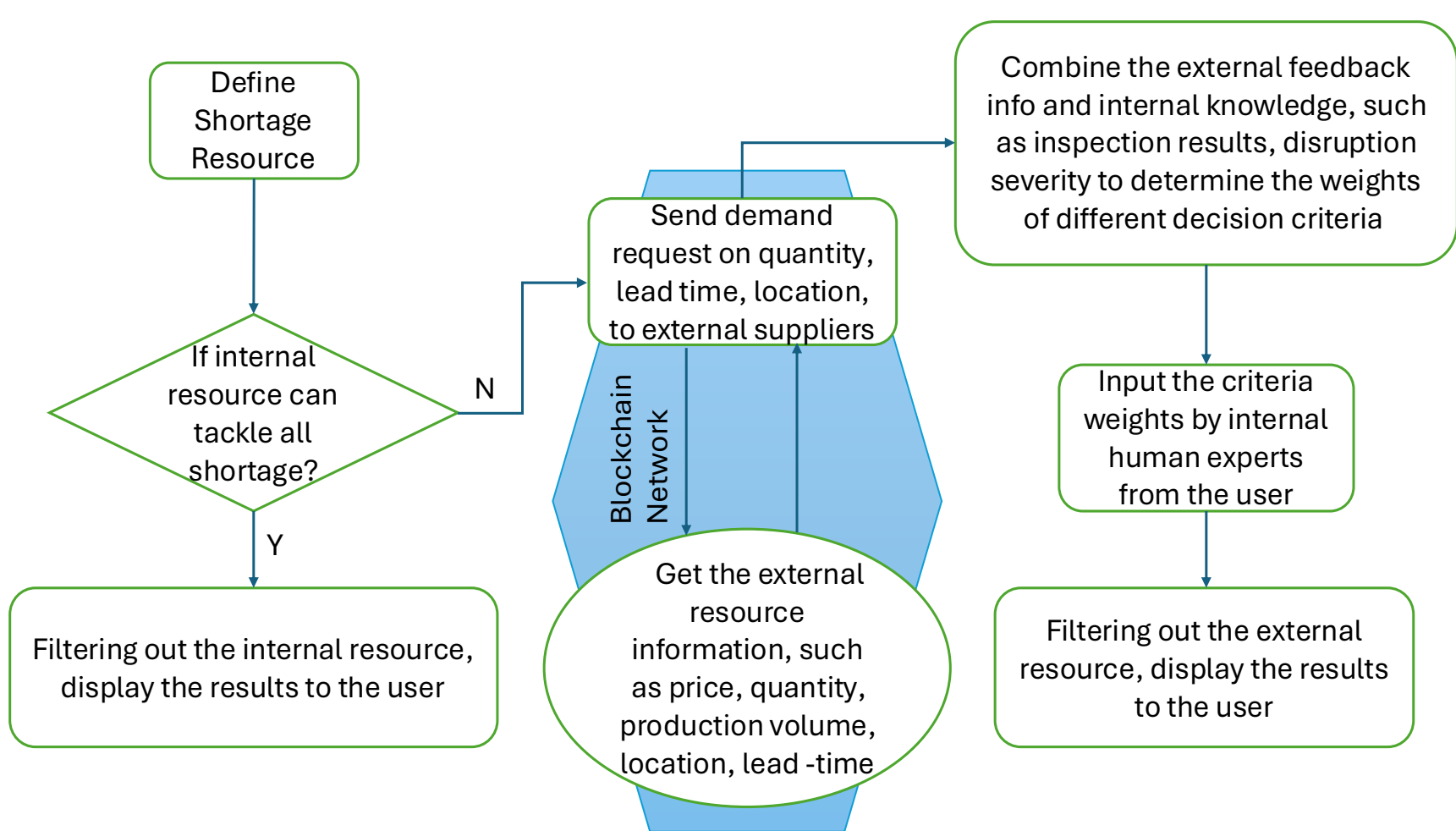

Figure 9. Decision process diagram of the BLC-IRS.

The proposed BLC-IRS system can optimize multiple objectives (Pachot et al., 2021a) while considering the resources within a SC network, including capacity and inventory space, truck-load utilization (Dadouchi and Agard 2021), optimal transit routes (Wang et al. 2014), and human resources (Hargaden and Ryan 2015). The system we proposed aims to react swiftly to SC disruptions. The external resource detection and recommendation function is tested in next section with an industrial use case.

### 3.6 BLC-IRS Framework Implementation

The developed BLC-IRS system framework can be utilized as an intelligent SC information system that aims to foster flexibility and reaction agility owned by SC participants.

The primary user of this BLC-IRS is the disrupted entity. The user can leverage this intelligent information system to determine the internal resource redundancy as a first response to SC disruption. Afterwards, it can also identify accessible redundant external resource from other partners within the same supply network. The IRS can promote results based on users' needs for the external resource rapidly, considering the practical limitations such as lead time, production capacity, pricing, and inspection scores. In this section this study illustrates a use case of the external resource recommendation, which is the essential function for reaching SCRes.

The dataset used to sketch external resource profiles and user profiles was sourced from the open data platform Kaggle at https://www.kaggle.com/datasets/harshsingh2209/supply-chain-analysis. The dataset covers most of the information we need for analysis and functionality demonstrating as most of the business supply chain data is confidential with limited access. User and resource profiles are illustrated in Table 3.

Table 3: User and Resource Profiles.

| User Profile | External Resources Profile |
|---|---|
| Product Type | Product Type |
| SKU | SKU |
| Availability | Price |
| Number of products sold | Supplier name |
| Revenue generated | Location |
| Stock levels | Lead time |
| Order quantities | Production volumes |
| | Inspection results |
| | Defect rates |
| | Transportation modes |
| | Routes |
| | Costs |

The original dataset comprising 100 observations and 24 variables, we assigned 17 nonduplicate features that can represent the profiles of the disrupted entity and the external resource owner. Inventory status (stock levels, availability), order information (order quantities) and sales information (number of products sold, revenue generated) are selected for establishing user's profile. And the external resource owner's profile is composed by the operational performance (production volumes, lead time), quality and reliability indicators (inspection results, defect rates), and logistics characteristics (transportation modes, routes and costs).

On the user side, the stock levels and availability data is the result after SC disruption. The inventory status after disruption serves as the main cause for emergency order placement to mitigate the shortage problem. Order quantity is essential information for communicating with the external supplier for emergency order placement. After the SC disruption, the BLC-IRS can detect user demand by analyzing features such as stock levels, internal availability, and order quantities.

The external resources profile will be composed of the features that were registered by the external resource owners, that is not confidential, such as the 'supplier name', location', 'lead time', emergency cost' (Cavalcante et al. 2019), and 'inspection results'. After the resource profile is generated, the knowledge-based recommendation algorithm will be selected since the weights of the recommendation criteria should be set beforehand based on internal experience or expert knowledge of the disrupted entity. The priority of each criteria will be set based on the disruption scenario, for example, in a very urgent case, the 'lead time' can be set at 0.6, along with other criteria cost, production volume, and inspection volume at different weights, to generate the recommendation results, and finally, multi-criteria recommendation can be applied to filter out the results that satisfied user's preference.

Supervised machine learning algorithms can be deployed when it is necessary to deal with well-patterned historical data for upcoming external resource prediction. Unsupervised machine learning algorithms or embedded deep learning can be employed to detect the data patterns and cluster the unpatterned data first when the input data from the external source is unorganized. This is to help the users get a quick overview of the external resources that are available. On the basis of these preliminary findings, knowledge-based recommendations can subsequently be applied to help the disrupted entity to locate accessible external resources information in chaos. Based on the procedure illustrated above, recommendation results for the case study are presented in Table 4, the dataset is static, as the dataset is intended to represent the post-disruption state of the user, and available external resource owner, the recommendation results will not be impacted without the explicit time stamps, real-time recommendation is not implemented.

Table 4: External Available Resource Recommendation Results.

| Demand | Recommended Resource | Overall score |
|---|---|---|
| SKU32 | Supplier 3 | 1.043993169 |

| SKU9 | Supplier 2 | 1.037888232 |
|---|---|---|
| SKU76 | Supplier 2 | 0.98991082 |
| SKU75 | Supplier 1 | 0.984446383 |

The overall score shows the performance ratings of external resource owners. The score is determined by the multi-criteria recommendation with priority weights of 0.35 for 'Lead time', 0.25 for 'Production volumes', 0.2 for 'Inspection results', and 0.2 for 'Costs'. These numerical weights are assigned based on the empirical data randomly and represent one single disruption response scenario that shorter leadtime is critical, to demonstrate the calculating function of the BLC-IRS. In the practical case, if the user adopts the BLC-IRS, the weights data will be determined by the internal human expert of the disrupted entity depending on their specific needs and experience. For example, in less time-critical scenarios, the weights of the 'Cost' can be considered with higher priority than the 'Leadtime'. Our present selection criterion indicates that the user identified Leadtime as the most important factor when choosing the appropriate outside resources. Rapidly identifying and utilizing available redundant external resource from other partners within the current supply network is the fundamental capability that allows the disrupted entity to mitigate the SC disruption before it finished the preparedness work for recovery. From the reactive side, the developed BLC-IRS can function as a novel and essential component of the SCRes strategy development.

# 4 BLC-IRS Framework Validation with Deployment Scope

## 4.1 BLC-IRS Effectiveness Validation with System Dynamics Simulation

The communication process of the disrupted entity after the SC disruption can be considered as two steps:

1) Inquiry the accessible external redundant resources and

2) Place emergency orders to replenish products in shortage after disruption.

When a business entity is disrupted, it will identify the most suitable external resource suppliers within the same supply network, communicate with them, place emergency orders as quickly as possible and wait for the replenishment of the physical product after the internal resources are exhausted.

The emergency order placement is the valid step that shifts the entire SC from a regular operation status to an emergency response operation status. During this process, the fundamental information and material flows after SC disruption are demonstrated in Figure 10.

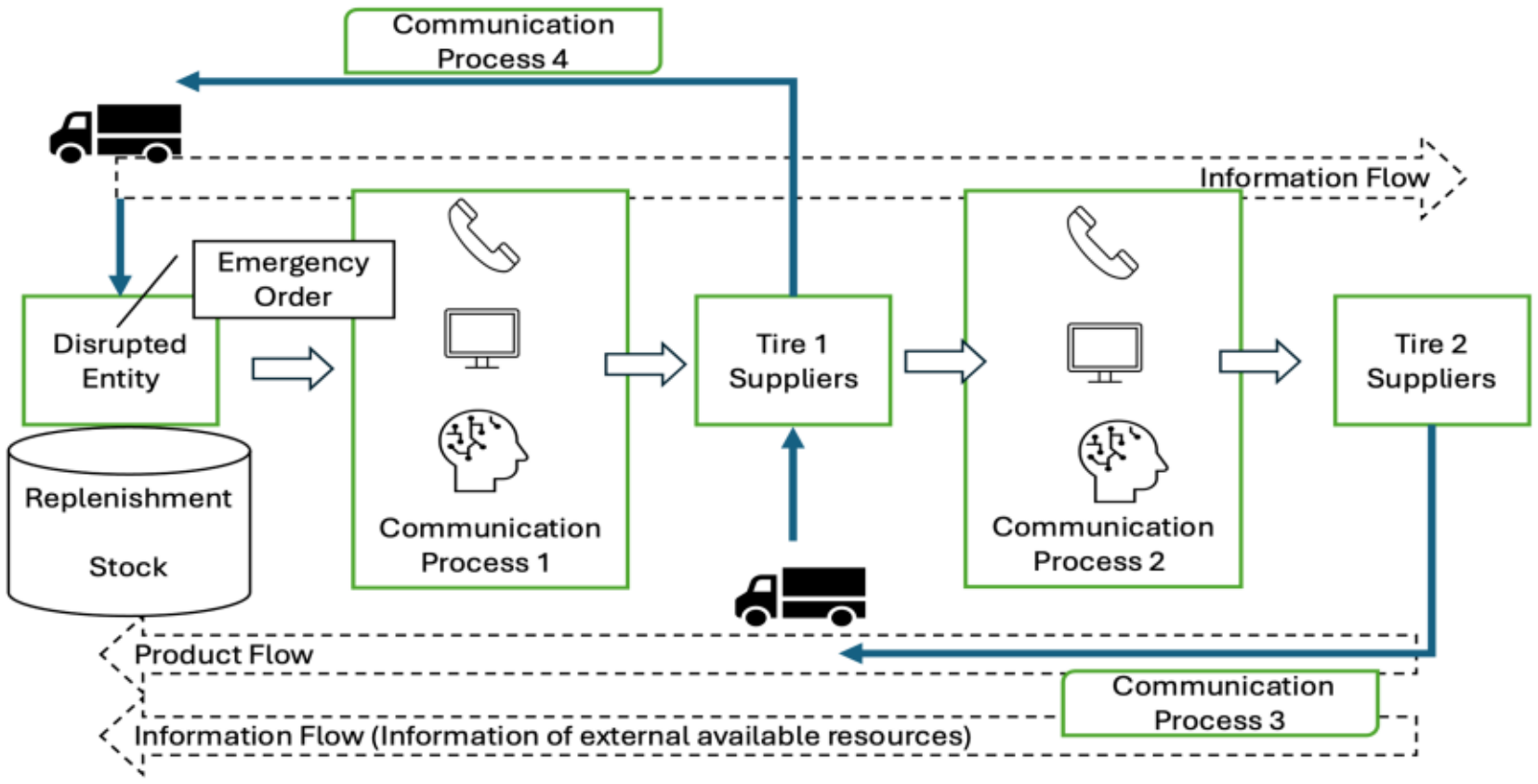

Figure 10: The movement of information and resources following supply chain interruption.

As the status of the entire SC will be changed to an emergency operation only after the order placement action of the disrupted entity. We can assume that the communication process formally ends with the order placement action after the order inquiry action. The effective communication process is composed by the order inquiry and order placement actions, therefore within a fixed communication timeframe, a reduced inquiry duration will lead to a higher frequency of order placements. For instance, by using traditional communication methods such as telephone or checking the fractional data exchange systems manually, users (the disrupted entity) may finalize 3-5 order placements within a designated time frame (60 minutes), as the inquiry procedure is comparatively time-consuming. In contrast, the proposed IRS enables the immediate presentation of accessible external resources to the user, hence accelerating the inquiry process, allowing order placement over 3-5 times within the same time frame.

In this case, the validation of the proposed BLC-IRS framework as an effective SCRes enhancement measure can be streamlined by examining the impact of modifying the order rate (Borshchev and Grigorey, 2020) with the same SC network condition. The KPI is to investigate the replenishment stock level of the with varying order rates to represent different communication scheme within a fixed duration, for example 50 days set in this study. In a fixed time, the higher replenishment stock level indicates a shorter lead time to meet the emergency demand of a fixed order quantity under identical conditions, which also suggests a higher SCRes level.

A System Dynamics (SD) model was developed to simulate a simple SC to validate the proposed BLC-IRS, as the SD approach is appropriate for assessing the efffectiveness of SCRes strategy (Olivares-Aguila and ElMaraghy, 2021), demonstrated in Figure 11. The rules and values pertaining to all parameters and variables were cited from Borshchev and Grigoryev (2020).

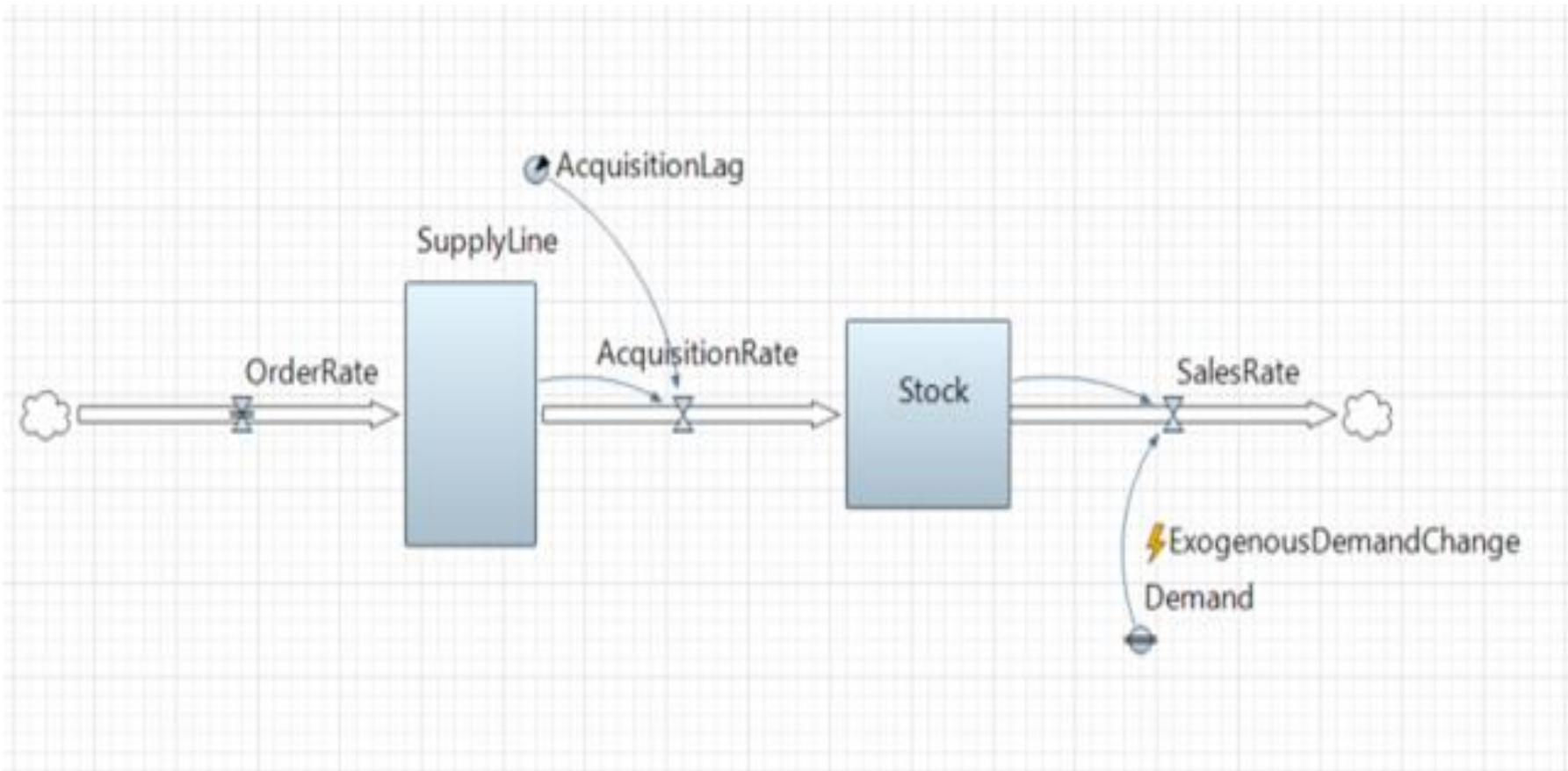

Figure 11: The system dynamics model to investigate the effects of the proposed IRS.

The 'SupplyLine' includes all supply activities involving external resources, including extra production of suppliers in different tiers of suppliers and transit activities by third-party logistics providers illustrated in Figure 8. The 'Stock' refers to the replenishment inventory of the user (disrupted entity). Both 'OrderRate' and 'Demand' are designated as constants; these two variables will be monitored externally from the SD model.

The demand will be altered by the 'ExogenousDemandChange' event, scheduled to transpire at each time unit (day). The action rule for 'ExogenousDemandChange' is defined as Demand = max (0, Demand + uniform (-1, 1)); this code modifies the Demand variable by a random number evenly distributed between -1 and 1, either increasing or decreasing it. The max () function prevents Demand from decreasing below zero. The 'SalesRate' uses the conditional operator Stock > 0? Demand: 0. When a product is available, it is sold at the demand rate, otherwise, no sales occur. In this experiment, only OrderRate is the variable that the user can monitor to evaluate the effectiveness of the proposed IRS. The experiment involves executing the model with varying OrderRates and examining the stock levels in a given time.

The model operates in step-pause mode, allowing the user to modify the order rate every 50 days. Table 5 reflects the difference in replenishment inventory level after SC disruption by varying the order rate. While the order placement frequency is set as 10, 4.976 and 12.217 with randomly to represent the traditional information sharing scheme, the OrderRate was set at 20, 6.951,13.553 to represent the BLC-IRS as the fundamental assumption was that the BLC-IRS can place more orders within the same time unit compared to the conventional communication methods.

Table 5 indicates that if we use the conventional information scheme, the arrived stock quantity will be lower than if we use the BLC-IRS after 50 days. Mathematically, in a fixed duration, the higher replenishment level shares the same meaning as less lead time in a fixed acquired stock quantity. The simulation results in Table 5 represent the same meaning as the results in Table 6.

Table 5: Comparison of stock levels of different communication schemes after SC disruption.

| Fixed Duration 50 Days | Order rate | 10(Conventional) | 20(BLC-IRS) |
|---|---|---|---|
| | Replenishment stock | 530.5 | 960.11 |
| | Order rate | 4.976(Conventional) | 6.951(BLC-IRS) |
| | Replenishment stock | 314.008 | 398.932 |
| | Order rate | 12.217(Conventional) | 13.553 |
| | Replenishment stock | 625.396 | 682.012 |

The results in Table 5 can be converted to the results in Table 6 when other conditions remain the same:

Table 6: Comparison of response time of different communication schemes after SC disruption.

| Fixed Quantity 530.5 Units | Communication Scheme | Conventional | BLC-IRS |
|---|---|---|---|
| | Fulfill Leadtime | 50 Days | 28 Days |
| 314.008 Units | | 50Days | 39Days |
| 625.39 Units | | 50 Days | 46Days |

These results in Table 5 can also be interpreted as if the disrupted entity acquired the arriving stock at 530.5 after disruption, and under the conventional communication scheme, the lead time will be 50 days, with the BLC-IRS, the lead time will be 28 days (Table 6). Based on the definition of SCRes, a shorter lead time to fulfilling the acquiring quantity after disruption means a higher resiliency level. The simulation results validated that our BLC-IRS can foster SCRes after disruption as a new communication tool.

### 4.2 BLC-IRS Deployment Scope

This proposed BLC-IRS framework can be applied to mitigate the SC disruption at firm level (node level) and partial network level (arch level) in physical SC network under the condition that SC participants connected well with each other in information system network. The BLC-IRS will be invalid under the whole SC network failure. The application scope is illustrated from Figure 12-14.

Figure 12 demonstrated that our proposed BLC-IRS can be applied and function as the SCRes enhancement tool at the firm (node) level disruption. Based on the complete implementation process from Subsection 3.5.3 to 3.6 the disrupted entity can detect available external resources, with the proposed BLC-IRS within the information network and get the replenishment from the most suitable external suppliers.

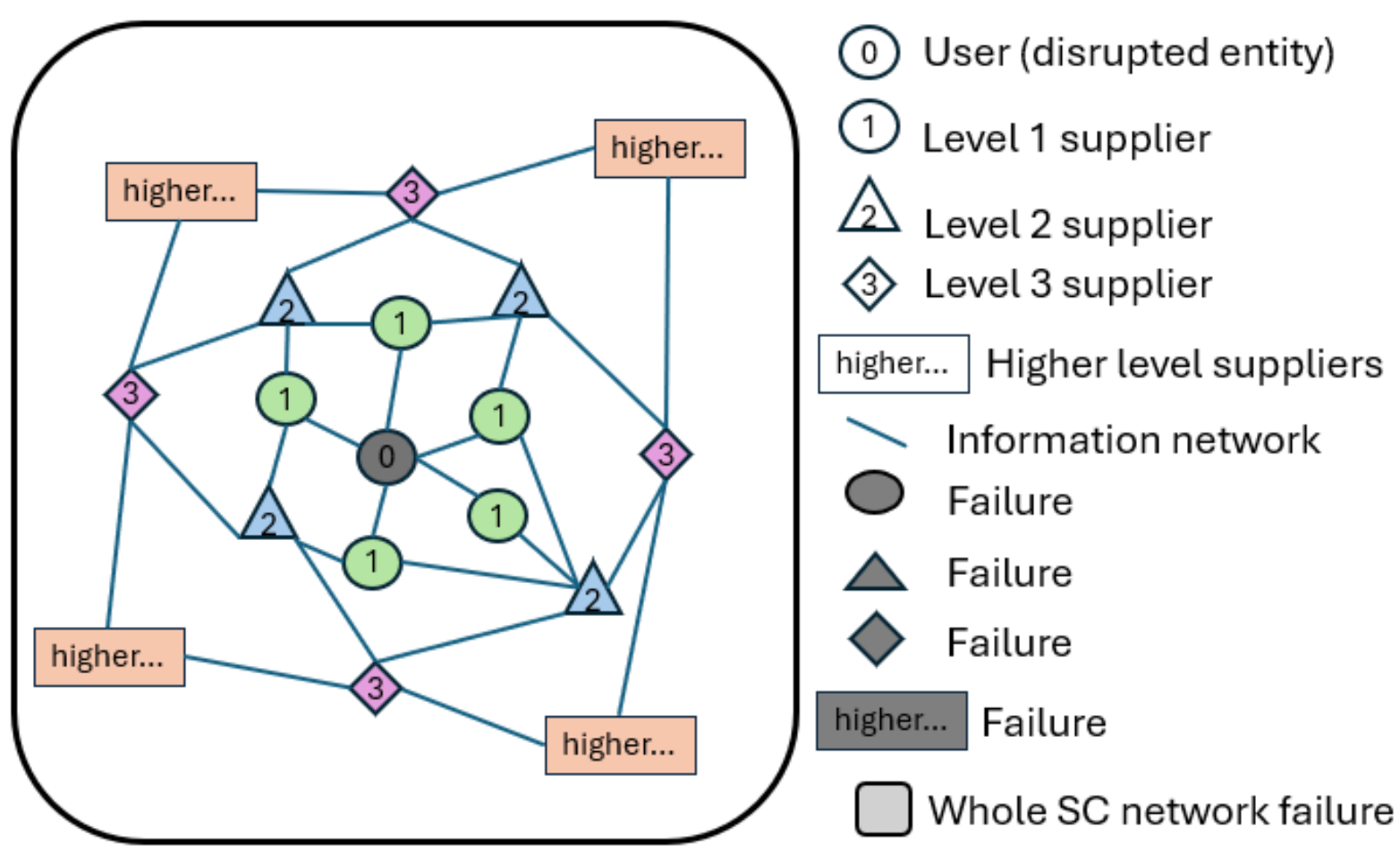

Figure 12: BLC-IRS adoption for firm level disruption in SC.

For the partial network failure illustrated in Figure 13, the proposed BLC-IRS can also be applied to assist the disrupted entity to find the most suitable existing available external resources with the same implementation logic described in Subsection 3.5.3 to 3.6.

In the partial network failure scenario, for the single disrupted entity 0, it can also find the most suitable external suppliers within the existing active supply channels with the proposed BLC-IRS. Besides the single disrupted entity 0, every other single participant within the partial failure SC network can use our proposed BLC-IRS to find the available external suppliers with the same logic and decision process if there are existing active supply channels. After all the single failure points finding available external resources with active supply channels to mitigate shortage through the BLC-IRS, we can consider that the local SC network level disruption is mitigated by the proposed BLC-IRS.

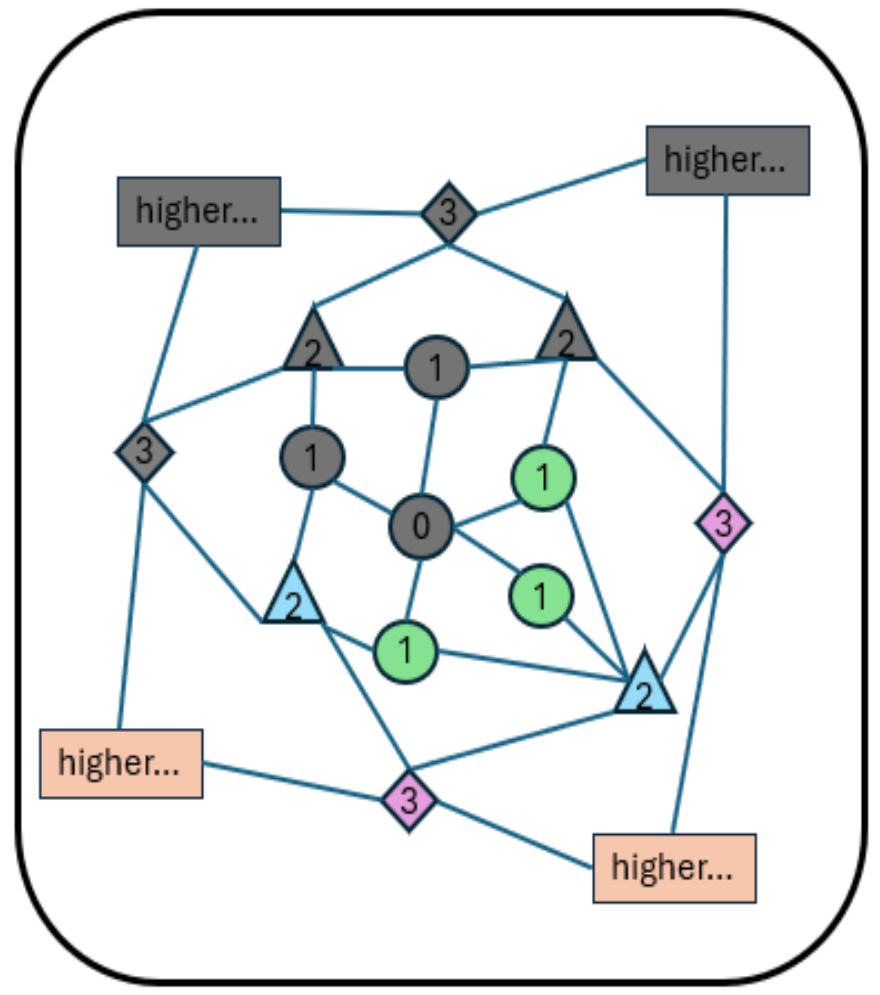

Figure 13: BLC-IRS adoption for partial network level SC disruption.

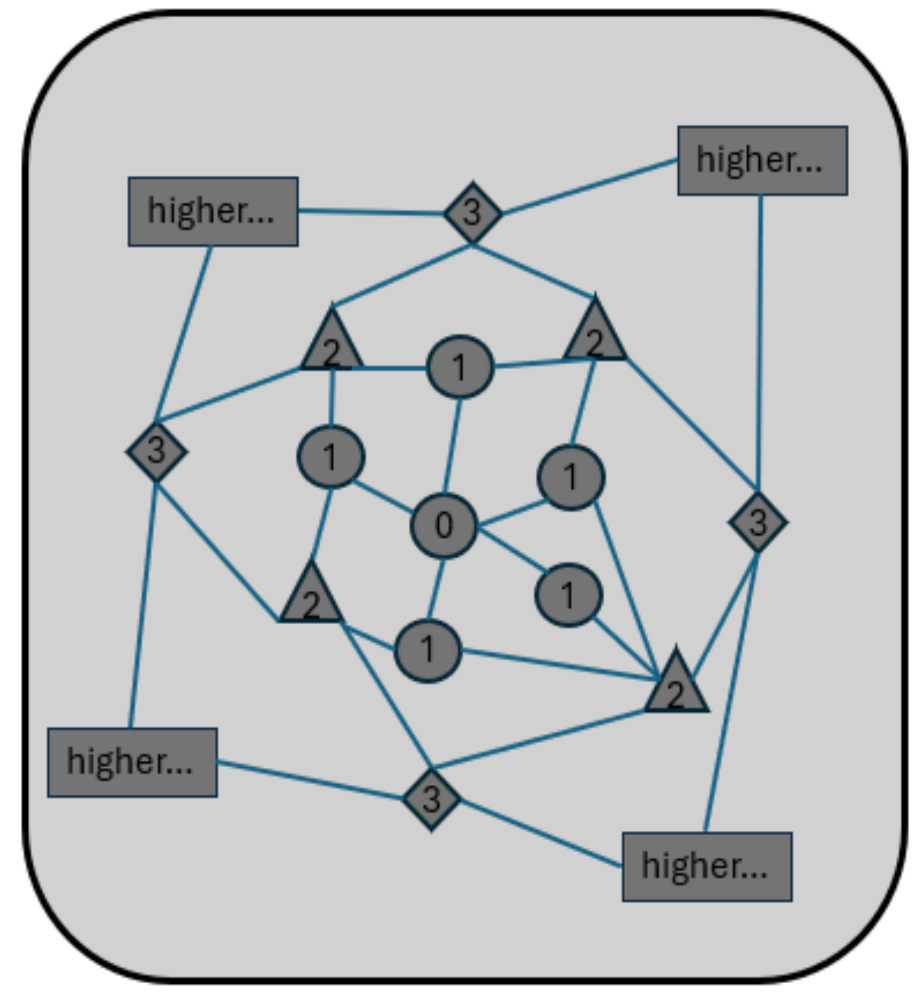

Figure 14: BLC-IRS adoption for whole network level SC disruption.

For the whole SC network level disruption illustrated Figure 14, the BLC-IRS will be infeasible to functioned as the SCRes enhancement tool in the response phase but can work in the recovery phase. In this scenario, even though the information network is alive, there is no active suppliers that can produce demanded materials to mitigate shortage after SC disruptions in the physical level for any failure dot, the shortage problem faced by all of the failure dots including the disrupted entity 0. The systematic disruption is unlikely to be mitigated by the BLC-IRS in the response phase in this case. After a period of time, the whole SC network failure in Figure 14 will evolve to the partial network failure illustrated in Figure 13, as the failure entities will find the long term solutions to back to the normal, such as rebuild production capacity by itself to mitigate shortage problem as there are no available external suppliers alive within the whole SC network. For example, during the early stage of COIVD-19, the whole global personal protect equipment (PPE) SC network was disrupted and faced the shortage problem, the vehicle maker company BYD transferred its car production line to PPE production and becomes the active PPE supplier to accelerate the speed to response to the PPE shortage. Besides enough PPE for its internal use, this company also becomes the one of the world's largest active PPE suppliers (Li et al., 2022b) that can activate other failure points connected with BYD and transferred the PPE SC network status from the whole network level disruption in Figure 14 to a partial network level disruption in Figure 13. Once the disruption level becomes the partial network level, the BLC-IRS can be adopted to accelerate the response and recovery speed for disruption mitigation. Table 7 summarizes the BLC-IRS implementation scope based on the illustration and discussion above.

Table 7： The BLC-IRS deployment scope based on the disruption level and reaction stage.

| Disruption Level | Response stage | Recovery stage |
|---|---|---|
| Firm (Node Failure） | BLC-IRS adoption ☑ | BLC-IRS adoption ☑ |
| Partial SC network (Arch Failure) | BLC-IRS adoption ☑ | BLC-IRS adoption ☑ |
| Whole SC network | BLC-IRS adoption ☒ | BLC-IRS adoption ☑ |

## 5 Discussion, Theoretical Contributions and Managerial Insights

This research contributes to the supply chain resilience literature by advancing disruption response research from predominantly conceptual and proactive discussions to an executable, reactive decision-support framework. Existing studies on digital technologies and SCRes have largely conceptualized blockchain or intelligent systems as standalone, proactive enablers of resilience, with limited attention to their operational integration or real-time applicability in disruption response scenarios. This research theorizes and demonstrates a synthetic and integrated use of blockchain and intelligent recommender systems, positioning such integration as a reactive SCRes mechanism rather than merely a strategic or preventive one.

Specifically, the proposed BLC-IRS framework contributes beyond prior conceptual research by (i) operationalizing SCRes theory into an executable system architecture, (ii) theoretically articulating how the complementary characteristics of blockchain (trust, immutability, decentralization) and intelligent recommender systems (adaptive decision-making under uncertainty) jointly address disruption response challenges, and (iii) establishing a structured linkage between SCRes theoretical constructs and concrete system design choices.

This research extends the application boundaries of blockchain and intelligent recommender systems within supply chain research, also theoretically substantiates the necessity and flexibility of hybrid digital tool integration as a distinct pathway for enhancing supply chain resilience, thereby bridging the gap between SCRes theory and actionable digital infrastructure design.

The preliminary conditions for the successful implementation of our proposed BLC-IRC as a valid SCRes enhancement measure are the high level of information collaboration (Tang, 2006a). For the practitioners, this research highlighted that information collaboration is an important SCRes antecedent (Scholten and Schilder, 2015). This has been evident after implementing the smart contract and the complete BLC-IRS framework with the practical use case. The delivery of this research is an executable tool at the operational level that provides empirical evidence for the latest tactical finding that 'firms using supply chain planning systems gain higher resilience levels' (Swink et al., 2025). Moreover, the simulation results indicated that a faster information processing speed will be beneficial for enhancing SCRes. For practitioners in general, developing the information exchange speed through appropriate measures represents an important SCRes strategy to be borne in mind.

## 6 Conclusion and Future work

This research presented a blockchain-based data-driven intelligent recommender system framework to mitigate supply chain disruption risk as an agile disruption response tool. The proposed BLC-IRS system can provide internal and external resource recommendations rapidly with the considerations on the practical constraints complying with the SC disruption, performed as reactive supply chain resilience enhancement measure because of its rapid response capability. The suggested BLC-IRS framework was demonstrated with a practical case application and validated by SD simulation as a new communication scheme after SC disruption. The simulation results indicated that our BLC-IRS can serve as an effective strategy for mitigating SC disruption during the response phase, enabling supply chain participants to react more effectively to disruptions, finally at the physical product level. The delivery of this research is adaptable for firm level (node level) and partial network level (arch level) SC disruption in the response and recovery phase, the BLC-IRS can also be deployed in the recovery phase after the whole SC network disruption.

The simulation outcomes can also provide a research basis for the future development of high-efficiency recommendation algorithms, as the surging data volume in future cases can reduce the response speed of the BLC-IRS as high efficiency disruption mitigation tool. This study demonstrated the BLC-IRS framework by using static datasets and only tested the externally available capacity recommendation scenario. Dynamic and real-time recommendation experiments can be conducted with concrete examples and use cases across various industrial sectors such as medical device or pharmacy SC (Dai and Tang, 2020) that related to not only economy but also social welfare. Different recommendation algorithms, including AI-based methods, can be employed. Different external resource recommendation scenarios, such as external inventory redundancy, truck-load optimization (Dadouchi and Agard 2021), optimal transportations routes (Wang et al. 2014), and human resource (Hargaden and Ryan 2015) recommendations can be validated to examine the scalability of the proposed framework.

In our current research, the weights of the recommendation criteria are fully determined by human experts under conditions of low disruption frequency. For organizations that possess a higher volume of historical disruption records, incorporating a mechanism for optimizing recommendation weights would be beneficial in future work.

Moreover, the recommendation result generated by this proposed BLC-IRS is based solely on information matching between the disrupted entity and the external supplier on both sides. Other participants or entities, such as third-party logistics providers and road transport conditions, that may affect the final delivery time, can be considered as separate peers in the blockchain network. At the same time, their information can be included in the information matching and recommendation process in the future to obtain more practical results. An essential element of this study is the successful collaboration attained, especially regarding information exchange among various supply chain partners, which can potentially pose a challenge in practical contexts (Cui et al., 2024). And for the small and medium enterprises (SMEs), the BLC-IRS may be challenging to implement, the additional decentralized BLC-IRS systems and services will lead to the IRS project running significantly behind schedule and massively over budget which will be especially a critical barrier to SMEs, it will be meaningful in the future to study how to build a universal benefit mechanism that can serve as an accessible digital (Tang, 2022) SCRes tool for SMEs to accomplish the SC fairness (Chen et al., 2022).